\documentclass[fleqn,usenatbib]{mnras}

\usepackage{newtxtext,newtxmath}

\usepackage[T1]{fontenc}

\DeclareRobustCommand{\VAN}[3]{#2}
\let\VANthebibliography\thebibliography
\def\thebibliography{\DeclareRobustCommand{\VAN}[3]{##3}\VANthebibliography}

\usepackage{graphicx}	
\usepackage{amsmath}	
\usepackage{amssymb}	
\usepackage{tablefootnote}

\title[OB stars and associations in Auriga]{Mapping the distribution of OB stars and associations in Auriga}

\author[A. L. Quintana et al.]{
Alexis L. Quintana$^{1}$\thanks{E-mail: a.l.p.quintana.isasi@keele.ac.uk},
Nicholas J. Wright$^{1}$ and Robin D. Jeffries$^{1}$ \\
$^{1}$Astrophysics Group, Keele University, Keele ST5 5BG, UK\\
}

\date{Accepted 2023 April 17. Received 2023 April 14; in original form 2023 February 15}

\pubyear{2023}

\begin{document}
\label{firstpage}
\pagerange{\pageref{firstpage}--\pageref{lastpage}}
\maketitle

\begin{abstract}
OB associations are important probes of recent star formation and Galactic structure. In this study, we focus on the Auriga constellation, an important region of star formation due to its numerous young stars, star-forming regions and open clusters. We show using \textit{Gaia} data that its two previously documented OB associations, Aur OB1 and OB2, are too extended in proper motion and distance to be genuine associations, encouraging us to revisit the census of OB associations in Auriga with modern techniques. We identify 5617 candidate OB stars across the region using photometry, astrometry and our SED fitting code, grouping these into 5 high-confidence OB associations using HDBSCAN. Three of these are replacements to the historical pair of associations - Aur OB2 is divided between a foreground and a background association - while the other two associations are completely new. We connect these OB associations to the surrounding open clusters and star-forming regions, analyse them physically and kinematically, constraining their ages through a combination of 3D kinematic traceback, the position of their members in the HR diagram and their connection to clusters of known age. Four of these OB associations are expanding, with kinematic ages up to a few tens of Myr. Finally, we identify an age gradient in the region spanning several associations that coincides with the motion of the Perseus spiral arm over the last $\sim$20 Myr across the field of view. 
\end{abstract}




\section{Introduction}
\label{intro}
First defined by \citet{Amb1947}, OB associations are gravitationally unbound groups of young stars containing bright O- and B-type stars. They have sizes from a few tens of parsecs to a few hundred parsecs and total stellar mass of one thousand to several tens of thousands of solar masses \citep{Wright2020}. They are valuable tracers of the distribution of young stars, and have been used for such purposes for decades (see e.g. \citealt{Morg1953} and \citealt{Humphreys1978}). Most of the known OB associations are coincident with the Galactic spiral arms \citep{Wright2020,Wright2022}.

\citet{Bok} pointed out that low-density systems were prone to disruption by tidal forces from the Galaxy, therefore \citet{Amb1947} and \citet{Blauuw} assumed that OB associations should be expanding. In the \textit{clustered} model of star formation from \citet{Lada2003}, massive stars forming in embedded clusters disperse their parent molecular cloud by feedback, a process known as \textit{residual gas expulsion} \citep{Hills,Kroupa2001}. With the majority of the mass of the system in the form of gas, embedded clusters unable to survive as gravitationally bound open clusters will expand and disperse as unbound OB associations. The \textit{hierarchical} model of star formation, on the other hand, assumes that stars form over a range of densities, quickly decoupling from the gas in which they form. High-density clusters may survive as long-lived open clusters, while low-density groups will be gravitationally unbound from birth \citep{Krui2012}. In such a model, OB associations may form gravitationally unbound and not require residual gas expulsion. Although the reality probably lies between these two cases \citep{Wright2020}, recent data and modern techniques can provide the key to unveil the origins of OB associations.

Expansion signatures from OB associations could indeed help to support the \textit{clustered} model. Attempts to detect expansion in OB associations have had varied results, with early studies finding very little evidence for expansion (see e.g. \citealt{Wright2016, WrightMamajek, Ward}), while later studies had more success (see e.g. \citealt{Kounkel2018, CantatGaudin2019, Armstrong2020, Quintana}). Failures to detect clear expansion signatures in OB associations have occurred mostly in systems with historically-defined membership (based on the position on the sky), while more recent studies that defined OB associations and their membership using spatial and kinematic information have proven more successful.

The Auriga constellation contains two OB associations identified and catalogued by \citet{Roberts} and \citet{Humphreys1978}, as well as numerous young stars \citep{Gyu,Pandey}, star-forming regions \citep{Paladini2003,Mellinger,Anderson2015} and open clusters \citep{CantatGaudin}. The Auriga constellation should intercept both the local arm and the Perseus spiral arm though few studies have focused on Galactic longitudes between 140$^{\circ}$ and 180$^{\circ}$ \citep{Marco}. \citet{Negueruela2003} suggested the Auriga region is a less populated part of these spiral arms. 

Aur OB1 is located at a distance of 1.06 kpc \citep{Melnik2020}. It includes the open cluster NGC 1960 and the dark cloud LDN 1525 located at 1.2-1.3 kpc \citep{Straiz2010}, and is undergoing intense star formation \citep{Panja2021}. 

Aur OB2 is located at a distance of 2.42 kpc \citep{Melnik2020}. Its main features are the open clusters Stock 8, Alicante 11 and Alicante 12 \citep{Marco}. It was first thought that Aur OB2 extended between Stock 8 and NGC 1893, but recent studies have placed them at different distances, suggesting they may not all be part of the same system \citep{Negueruela2003,Marco, Kuhn}. 

The paper is structured as follows. In Section \ref{aurigaregion} we revisit the historical Auriga OB associations with modern data and techniques. In Section \ref{identification} we outline our process for identifying OB stars, before detailing the clustering process used to identify new OB associations. In Section \ref{analysis} we characterize these associations both physically and kinematically. In Section \ref{discussion} we discuss the results in a broader context and we provide conclusions in Section \ref{conc}.
\section{The Auriga region}
\label{aurigaregion}
In this section we explore the existing OB associations in Auriga with modern photometry and astrometry from \textit{Gaia} EDR3 \citep{GaiaEDR3}, as well as any known open clusters and star-forming regions in their vicinity. 
\subsection{Historical OB associations}
\label{newinv}

\begin{figure*}
    \centering
    \includegraphics[scale =0.45]{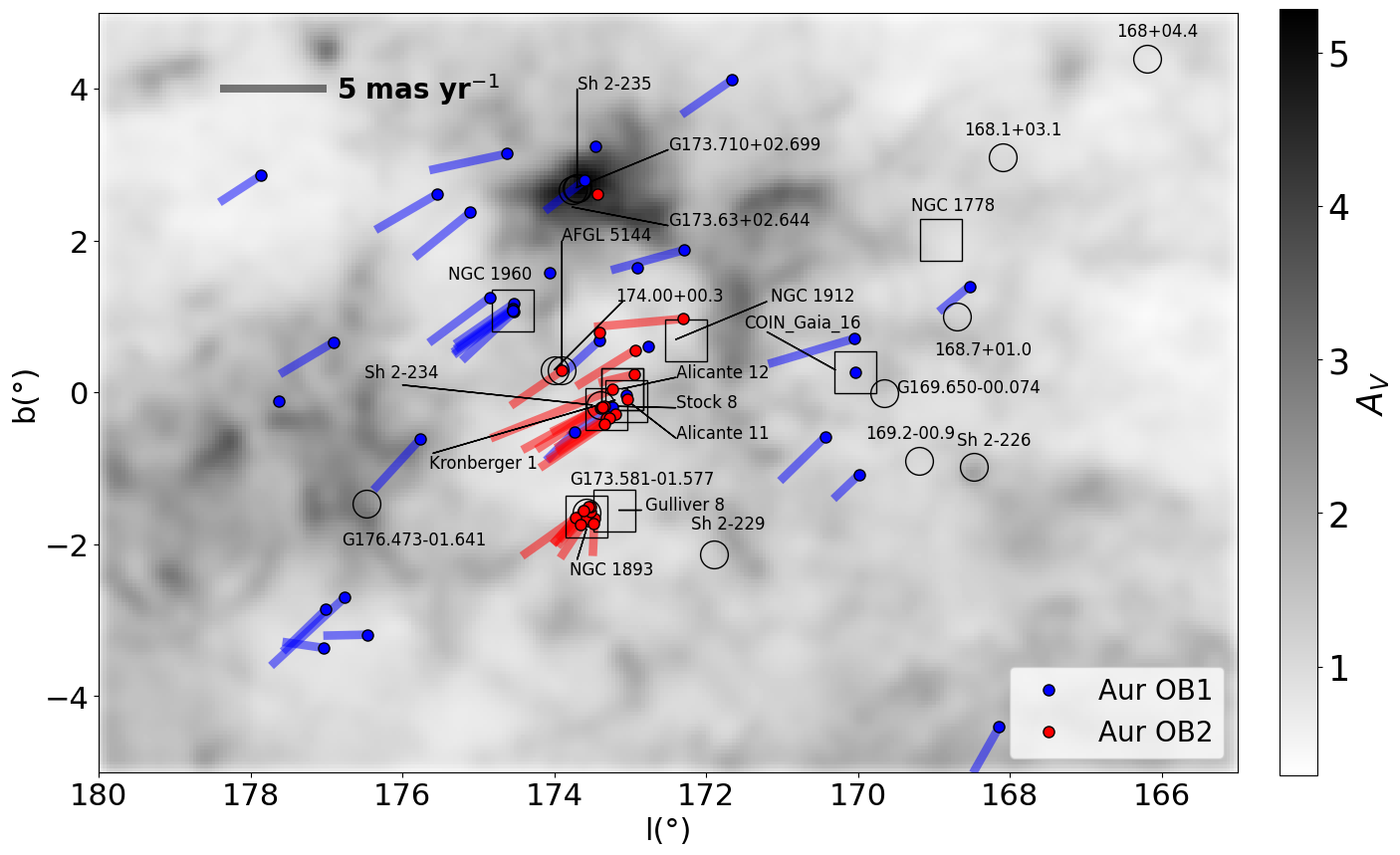}
    \caption{Spatial distribution in Galactic coordinates of the historical members of Aur OB1 and OB2. For the 48 stars with $RUWE < 1.4$, their Galactic proper motions are represented as vectors (scale length indicated in the top left) while the stars without reliable proper motions are shown as points. We also show open clusters as empty squares \citep{CantatGaudin}, and  H{\sc ii} and star-forming regions as empty circles \citep{Paladini2003,Mellinger,Anderson2015}. The background extinction map shows the integrated visual extinction at 2 kpc from \citet{Bayestar}.  \label{AurSpace}}
\end{figure*}

We focus our study on a 150 deg$^2$ area in the Auriga constellation, with $l$ = [165$^{\circ}$, 180$^{\circ}$] and $b$ = [-5$^{\circ}$, 5$^{\circ}$] as shown in Fig. \ref{AurSpace}. This area encompasses two historical associations, Aur OB1 and OB2. Their members have been listed in several catalogues (e.g. \citealt{Humphreys1978, Melnik2020}). From \citet{Melnik2020} there are 36 stars in Aur OB1, 20 in Aur OB2 and 10 in NGC 1893, although only 6 of them have equatorial coordinates in \textit{Gaia} EDR3 and listed in SIMBAD. NGC 1893 is usually considered part of Aur OB2 (see e.g. \citealt{Marco, Lim2018}), and we follow that convention here, increasing the number of Aur OB2 members to 26 stars. 

\begin{figure}
    \centering
    \includegraphics[width = \columnwidth]{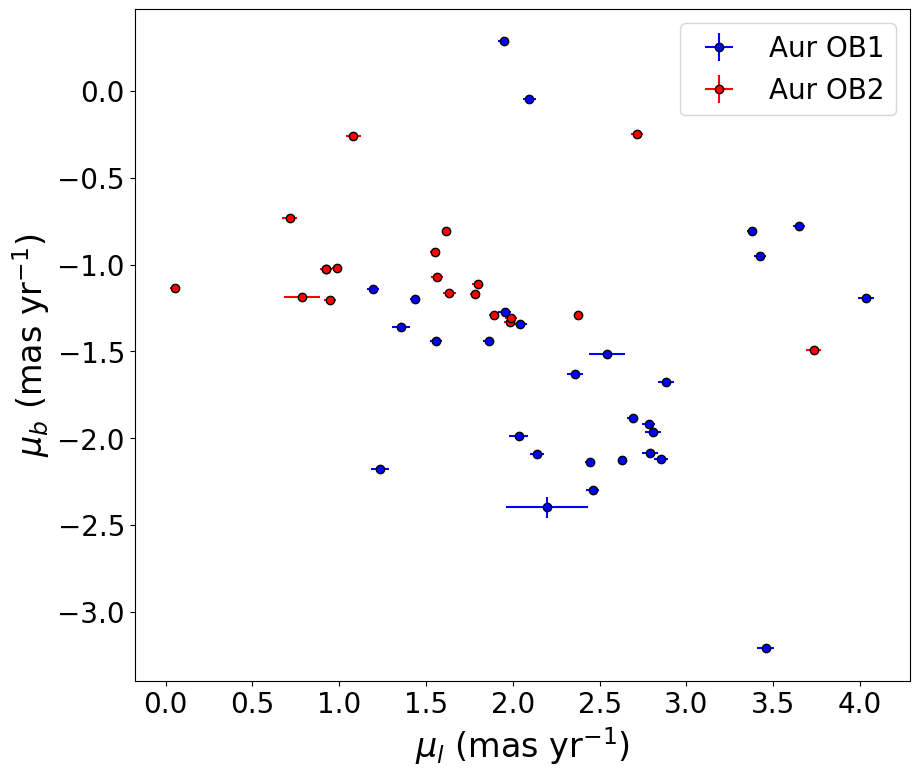}
    \caption{Proper motion distribution in Galactic coordinates for the historical members of Aur OB1 and OB2, with error bars, for stars with $RUWE < 1.4$. \label{AurPm}}
\end{figure}

\begin{figure*}
    \centering
    \includegraphics[scale = 0.5]{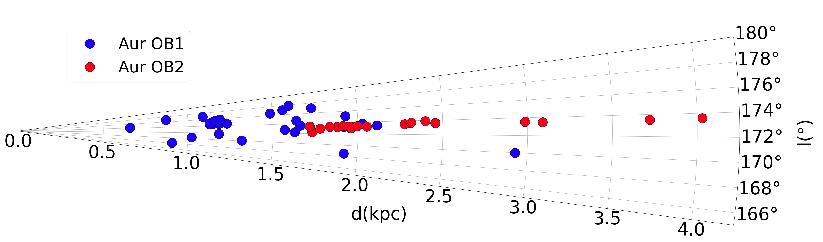}
    \caption{Galactic longitude plotted as a function of distances from \citet{BailerEDR3} for the historical members of Aur OB1 and OB2 with $RUWE < 1.4$. \label{AurDist}}
\end{figure*}

We match these 62 sources with \textit{Gaia} EDR3 \citep{GaiaEDR3} using a radius of 1'' and find a counterpart for all the stars. Following the criterion from \citet{Lindegren2020}, we only use the astrometry for the 48 stars whose renormalised united weight error (RUWE) is $<1.4$. Distances were taken from \citet{BailerEDR3}. The distribution of these stars in position, proper motions and distance is shown in Figs. \ref{AurSpace}, \ref{AurPm} and \ref{AurDist}.

Figures \ref{AurSpace} and \ref{AurPm} show that the existing members of the two associations do not have a strong level of kinematic coherence, their proper motions each spread over 2-3 mas yr$^{-1}$ or 10-15 km s$^{-1}$ at 1 kpc, much larger than one would expect for an OB association \citep{Wright2020}. Figure \ref{AurDist} shows that the Aur OB1 members are spread over distances from 0.6 to > 2 kpc, much larger than the parallax uncertainties (typically 0.03 mas). A similar issue is apparent for Aur OB2, its members are spread from 1.7 to over 4 kpc, albeit with a core group of stars around 2 kpc, though this does not match with the distance to NGC 1893 of 2.9 kpc \citep{Melnik2009, Melnik2020}. The presence of stars at distances of 3-4 kpc within these associations was previously noted by \citet{Marco}. OB associations have historically been defined through their on-sky spatial distribution and apparent magnitudes, with their members assumed to be within a narrow range of distances (see e.g. \citealt{Humphreys1978}). It is clear that these two associations are not real OB associations; they neither exhibit the necessary kinematic coherence, nor are they located at a small enough range of distances to have been born together.
\subsection{Open clusters and star-forming regions}
To revisit our census of the OB associations in Auriga, we start by collating information on the known open clusters and star-forming regions in this area. Several tens of open clusters have been identified in the region \citep{CantatGaudin}. In particular, five of them are likely related to the existing OB associations, following the discussions in \citet{Straiz2010}, \citet{Marco} and \citet{Pandey}. The properties of these OCs are summarized in Table \ref{AurOCs}\tablefootnote{Alicante 11 and 12 are not listed in \citet{CantatGaudin}. However, \citet{Marco} calculated a common distance of $\sim$ 2.8 kpc for these two clusters along with Stock 8, albeit overestimated compared with other estimates \citep{Jose2008,Melnik2009}, so we assigned them the same distance as Stock 8 in \citet{CantatGaudin}.}, where we have also included other OCs in the region whose relevance will be shown in Section \ref{Comphistoc}. The clusters are also shown in Fig. \ref{AurSpace} alongside the OB associations.

\begin{table}
	\centering
	\caption{Properties of the open clusters in Auriga thought to be related to the OB associations. Galactic coordinates and distances taken from \citet{CantatGaudin}. References for the ages are: \citet{Jeffries} and \citet{Joshi} for NGC 1960, \citet{Marco} for Stock 8, Alicante 11 and 12, \citet{Tapia}, \citet{Marco2001}, \citet{Sharma} and \citet{Lim2014} for NGC 1893, \citet{Sub1999}, \citet{Dias2021} for Gulliver 8, \citet{Jacobson2002}, \citet{Pandey2007}, \citet{Kharchenko2005} and \citet{Dib} for NGC 1912, \citet{Barbon1973}, \citet{Kharchenko2013}, \citet{Dib} and \citet{CantatGaudin2020b} for NGC 1778, \citet{CantatGaudin2020b} for COIN-Gaia\_16 (here abbreviated CG16), \citet{Dib} and \citet{CantatGaudin2020b} for Kronberger 1 (here abbreviated K1). \label{AurOCs}}
	\renewcommand{\arraystretch}{1.3} 
	\begin{tabular}{lccccccccr} 
		\hline
		OC & Assoc. & $l \, (\circ)$ & $b \, (\circ)$ & $d$ (kpc) & Age (Myr) \\
		\hline
		NGC 1960 & Aur OB1 & 174.542 & 1.075 & $1.16 \pm 0.01$ & 18-26 \\
		Stock 8 & Aur OB2 & 173.316 & -0.223 & $2.11 \pm 0.01$ & 4-6\\
		Alicante 11 & Aur OB2 & 173.046 & -0.119 & $2.11 \pm 0.01$  & 4-6 \\
		Alicante 12 & Aur OB2 & 173.107 & 0.046 & $2.11 \pm 0.01$ & 4-6 \\
		NGC 1893 & Aur OB2 & 173.577 & -1.634 & $3.37 \pm 0.05$ & 1-5 \\
        Gulliver 8 & - & 173.213 & -1.549 & $1.11 \pm 0.01$ & 22-39 \\
        NGC 1912 & - & 172.270 & 0.681 & $1.10 \pm 0.01$ & 250-375 \\   
        NGC 1778 & - & 168.914 & 2.007 & $1.64 \pm 0.01$ & 150-282 \\
        CG16 & - & 170.038 & 0.270 & $1.53^{+0.02}_{-0.01}$ & 26 \\
        K1 & - & 173.106 & 0.049 & $2.12 \pm 0.06$ & 6-8 \\
		\hline
	\end{tabular}
\end{table}
In this area are also found multiple H{\sc ii} regions \citep{Paladini2003,Anderson2015}, and several star-forming regions including Sh 2-235 and AFGL 5144 \citep{Mellinger}. They are shown in Fig. \ref{AurSpace}.

The most prominent feature of Fig. \ref{AurSpace} is the centre of the region at $l \sim 173^{\circ}$ and $b \sim 0^{\circ}$. This is where the bulk of Aur OB2 members are located \citep{Melnik2020}, along with the three open clusters Stock 8, Alicante 11 and 12 (see Table \ref{AurOCs}), and the H{\sc ii} regions Sh 2-234 and 174.0+00.3. The star-forming region AFGL 5144 lies close to this area, at $l = 173.7^{\circ}$ and $b = 0.3^{\circ}$ \citep{Mellinger}, consistent with the young age of the OCs \citep{Marco}.

The star-forming region Sh 2-235 is located at $l = 173.7^{\circ}$ and $b = 2.7^{\circ}$ \citep{Mellinger}, close to the H{\sc ii} regions G173.710+02.699 and G173.63+02.664, and where the region of highest extinction can be found (see Fig. \ref{AurSpace}).

\section{Identification of new OB associations}
\label{identification}
In this section we summarize the method used to identify OB stars and associations. The method for identifying OB stars is very similar to that of \citet{Quintana}, which we briefly summarise here and highlight any changes.
\subsection{Data and selection process}
\label{data}
We utilise astrometry and optical photometry from \textit{Gaia} EDR3 \citep{GaiaEDR3} \footnote{With the parallaxes corrected for the zero-point following the prescription from \citet{ZeroPoint}}, optical photometry from IGAPS\footnote{the INT Galactic Plane Survey} \citep{Drew,Mongui}, and near-IR photometry from 2MASS\footnote{2 Micron All Sky Survey} \citep{2MASS} and UKIDSS \footnote{United Kingdom Infrared Deep Sky Survey} \citep{UKIDSS}. We require \textit{Gaia} astrometry to have $RUWE < 1.4$ and $|\frac{\varpi}{\sigma_{\varpi}}| > 2$ \footnote{We have not applied the correction on parallax uncertainty from \citet{ElBaldry}, because our analysis of the line of sight distribution of OB stars within our new associations suggests that the \textit{Gaia} parallax uncertainties are overestimated in this area.}, where $\varpi$ is the observed \textit{Gaia} parallax and $\sigma_{\varpi}$ its random uncertainty. We limit our sample to stars with BP-RP < 2.5, a colour limit equivalent to a star with $\log(T_{\rm eff}) = 4$ and $A_V = 6$, which is about the maximum extinction level in this region at a distance of 3 kpc \citep{Bayestar}. The sources were filtered to have $d$ < 3.5 kpc, using the distances from \citet{BailerEDR3}. 

\textit{Gaia} photometry was required to have $|C^*| < 3 \, \sigma_{C*}$ where $C^*$ is the corrected excess flux factor in the $G_{\rm RP}$ and $G_{\rm BP}$ bands and $\sigma_{C*}$ is the power-law on the $G$ band with a chosen 3$\sigma$ level \citep{Riello}. 2MASS photometry was required to have a good quality flag (A, B, C or D, see \citealt{2MASS}) whilst those from UKIDSS had to fulfill $ErrBits < 256$. For UKIDSS we also exclude photometry with either $J < 13.25$, $H < 12.75$ and $K < 12$, below which the photometry risks saturation \citep{UKIDSS}. IGAPS photometry was filtered by excluding saturated photometric bands whose associated class did not indicate a star or probable star \citep{Mongui}. We then require at least one valid blue photometric band (either $g$, $G_{\rm BP}$ or $G$) and a valid near-infrared photometric band. 

To remove faint (non-OB) stars, we then apply an absolute magnitude cut, requiring $M_K < 1.07$ (if K-band photometry is available), $M_H < 1.10$ (otherwise if H-band photometry is available), or $M_J < 1.07$ (if only J-band photometry is available). These are the absolute magnitudes of main-sequence A0 stars \citep{Mamajek}. 

Finally, the near-IR colour-colour diagram was used to remove background giants, as described in \citet{Quintana}.

This led to a working sample of 29,124 sources on which we applied our SED fitting process.

\subsection{SED fitting}
\label{SED}
To calculate the physical properties of the sources, in order to identify OB stars, an SED fitting process was applied, based on the same method in \citet{Quintana} with a few improvements, summarised here:
\begin{itemize}
\item We seek to estimate the model parameters log(Mass), Fr(Age), $d$ and $\ln(f)$ using the \textit{emcee} package in Python \citep{Emcee}. Fr(Age) is the fractional age (i.e. the age of a star divided by the maximum age at its initial stellar mass) and $\ln(f)$ is a scaling uncertainty to help the convergence of $\chi^2$ \citep{Emcee, Casey}. $\log(T_{\rm eff})$ and $\log (L/L_\odot)$ are indirect products of this process, and the extinction $A_V$ was derived using the 3D extinction map from \citet{Bayestar} named \textit{Bayestar}. The priors for these parameters are:
\begin{equation}
\label{priors}
\ln (P(\theta)) =
\begin{cases}
\log (\frac{1}{2 \, L^3} \, d^2 \, \exp{(\frac{-d}{L}})) & \text{if }
\begin{cases}
-1.0 \leq \log(\rm Mass) \leq 2.0  \\
0.0 \leq \rm Fr( Age)  \leq 1.0 \\
0.0 \leq d \leq 5000.0 \, \rm pc \\
-10.0 \leq \ln(f) \leq 1.0 \\
\end{cases} \\ 
- \infty  & \text{otherwise}
\end{cases}
\end{equation}
\noindent with the prior on distance from \citet{Bailer2015} including a scale length $L$ set to 1.35 kpc.
\item Our model SEDs use stellar spectral models \citep{Werner1999,Rauch, Werner, Coelho}, with a fixed value of $\log g = 4 $ and evolutionary models from \citet{Ekstrom}. Model spectra were reddened using the \citet{Fitzpatrick} extinction laws and convolved with the relevant filter profiles to derive synthetic magnitudes.
\item Systematic uncertainties were added to the measured photometric uncertainties. This is equal to 0.03 mag for $g$, $r$ and $i$ \citep{BarentsenArt, Drew2014}, 0.01 mag for $G$, $G_{\rm RP}$, $G_{\rm BP}$ \citep{Riello}, 0.03 mag for $J_{\rm 2M}$, 0.02 for $H_{\rm 2M}$ and $K_{\rm 2M}$ \citep{Skrutsie}, and 0.03 mag for $J_{\rm U}$, $H_{\rm U}$, $K_{\rm U}$ \citep{Hodgkin}.
\item We choose the median value of the posterior distribution. The posterior distribution was explored using a Markov Chain Monte Carlo simulation. This utilised 1000 walkers, 200 burn-in iterations and 200 iterations. If the $\ln(f)$ value was greater than 4, or the difference between the 95th and 5th percentile of $\log(T_{\rm eff})$ was greater than 0.5 (indicating a lack of convergence), we ran 1000 supplementary burn-in and 200 supplementary iterations, until convergence was achieved or for 6000 supplementary burn-in iterations.
\end{itemize}
In addition, the extinctions from \textit{Gaia} DR3 \citep{Creevey, Delchambre} reveal that the \textit{Bayestar} extinctions tend to be underestimated by $\sim$22 \%. Instead of using  the \textit{Gaia} DR3 extinctions (due to their incomplete coverage of the Galactic plane, \citealt{Delchambre}), we increase the \textit{Bayestar} extinctions by 22 per cent to compensate.
\subsection{General results}
\label{genresults}
\begin{figure*}
    \centering
    \includegraphics[scale = 0.09]{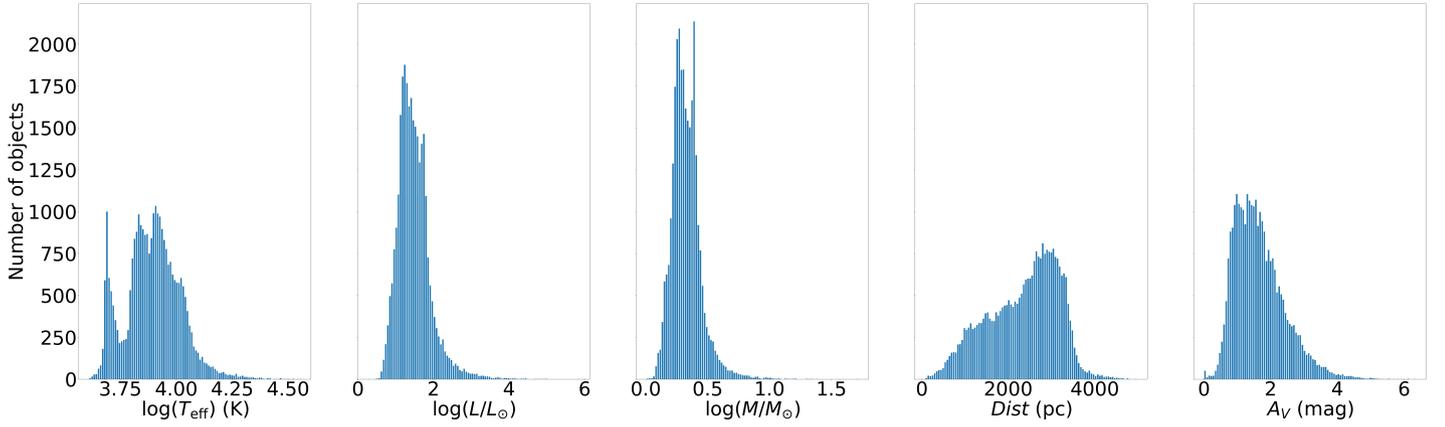}
    \caption{Median fitted parameters for the 29,124 selected sources of the working sample. \label{Fittedparam}}
\end{figure*}

SED fits were performed for all 29,124 candidate OB stars. Histograms of fitted physical parameters are shown in Fig. \ref{Fittedparam}. There are 5434 stars with $\log(T_{\rm eff}) > 4$ (OB stars, 18.66 \%) and 115 stars with $\log(T_{\rm eff}) > 4.3$ (O stars, 0.39 \%). The median value of $\log(M / M_{\odot})$ is equal to 0.31 (with a standard deviation of 0.12 dex) while the median value of $\log(L / L_{\odot})$ is 1.43 (with a standard deviation of 0.44 dex). Most of the stars are located within 4 kpc (consistent with our selection from \citealt{BailerEDR3}) with an increasing number at larger distances (as we probe a larger volume), while the peak of reddening is located at 1.5 mag, with the bulk at $A_V < 3$ mag.
\subsection{Incompleteness}
\begin{figure}
    \centering
    \includegraphics[width = \columnwidth]{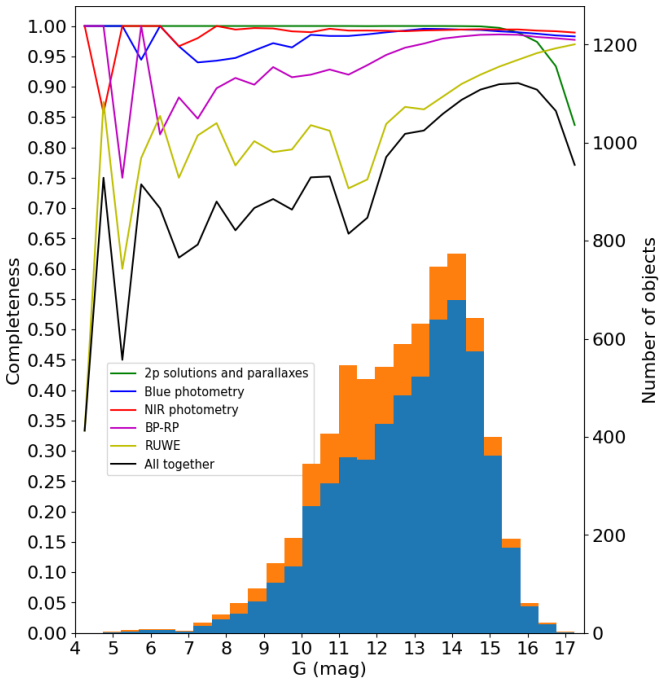}
    \caption{Completeness as a function of $G$ for the 5617 SED-fitted OB stars in the sample divided according to the different steps used to trim the sample. The black curve represents the product of all completeness curves. The blue and orange histograms show the number of sources before (blue) and after (orange) the completeness correction is applied.    \label{Incompleteness}}
\end{figure}

Incompleteness in the working sample stems from the selection process. To estimate it, we compute the fraction of stars as a function of magnitude which were trimmed during the successive steps of Section \ref{data}. These steps include the removal of bad astrometric solutions (2-parameter sources, large error on parallaxes and large $RUWE$), the removal of bad photometry (blue or NIR) and high BP-RP values. A plot of the completeness level as a function of $G$ is shown in Fig. \ref{Incompleteness} for the SED-fitted OB stars (stars with $\log(T_{\rm eff}) > 4$ or $\log (L/L_\odot) > 2.5$). 

To further verify the completeness of our sample, we crossmatch it with the OBA stars from \citet{Zari2021}. Their list contains 14,973 stars in the Auriga region and from the 29,124 stars in our sample, there are 4818 stars in common, including 4097 with a SED-fitted $T_{\rm eff}$ greater than 8000 K (the minimum temperature for \citealt{Zari2021}). Unsuccessful matches for our list are due to a different $M_K$ threshold (we chose $M_K$ < 1.07 while they selected stars with $M_K$ < 0). Unsuccessful matches from their list are due to our selection process (e.g. we discarded distant stars that they kept). As we estimated the incompleteness due to our selection process (Fig. \ref{Incompleteness}), this comparison shows that we have reached good completeness in probing the population of OB stars in Auriga.
\subsection{Comparison with spectroscopic temperatures}
\begin{figure}
    \centering
    \includegraphics[scale = 0.3]{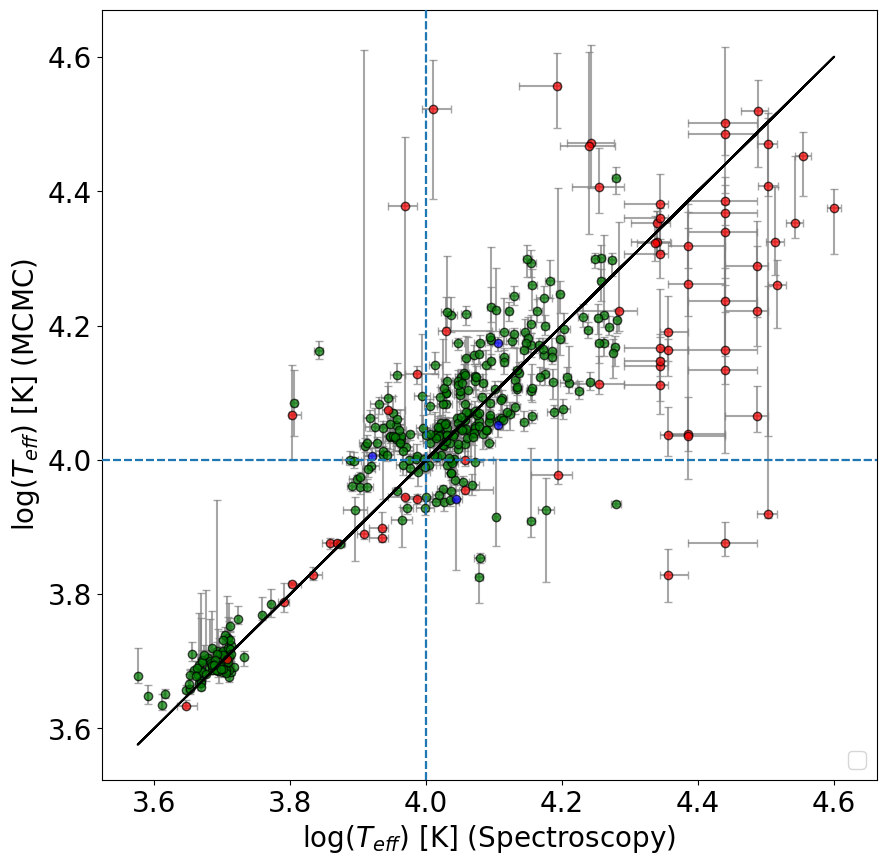}
    \caption{Comparison between the spectroscopic and SED-fitted temperatures for the 397 stars in the Auriga sample. Stars coloured in green are from APOGEE, stars coloured in red are from SIMBAD and stars coloured in blue are in both samples.}
    \label{TempAur}
\end{figure}

To check the quality of the results, we build a sample of spectroscopic temperatures that we compare to our SED-fitted temperatures, by cross-matching our sample within 1 arcsec with two catalogues:
\begin{itemize}
    \item Stars with spectral types from SIMBAD, filtered by removing sources with a quality measurement on spectral type of 'D' and 'E', along with those without an indicated spectral type and subclass. We then convert the spectral types into effective temperatures using the tabulations from  \citet{Martins} for the O-type stars (observed scale), from \citet{Trundle} for early B-type stars, from \citet{Humphreys1984} for late B-type stars of luminosity classes 'I' or 'III' and from \citet{Mamajek} for the later spectral types. We set a luminosity class of 'V' when unspecified and chose error bars of one spectral subclass, whilst using the spectral type of the primary star for binaries and interpolating for luminosity classes of 'II' and 'IV'.
    \item Stars from APOGEE DR17 \citep{Garcia,Abdu}, selecting the sources with a measured $T_{\rm eff}$ from the pipeline and removing those with a warning on $T_{\rm eff}$ that are considered unreliable due to their proximity to the upper limit of APOGEE measurements (20,000 K). 
\end{itemize}
We combine 70 stars from SIMBAD with 331 stars from APOGEE, making a sample of 397 unique stars (we use the weighted mean to calculate the temperature for the 4 stars in common). Our SED-fitted temperatures are compared with the spectroscopic temperatures in Fig. \ref{TempAur}.

Choosing thresholds on $\log(T_{\rm eff})$ > 4 (4.1, 4.2 and 4.3), we define the recovery rate ${\rm RR} = {\rm TP}/({\rm TP} + {\rm FN})$ where TP is the number of true positives (where both the SED-fitted and spectroscopic temperatures are above the threshold for selection of an OB star) and FN the number of false negatives (where the SED-fitted temperature is below the threshold and the spectroscopic temperature above). We also define the contamination rate, ${\rm CR} = {\rm FP}/({\rm TP} + {\rm FP})$, where FP is the number of false positives (where the SED-fitted temperature is above the threshold and the spectroscopic temperature below). For these thresholds, RR is equal to 88\% (80\%, 56\%, 49\% ) and CR is between 17 and 30 \%. These results suggest we are better at fitting late B-type stars, which could be due to the sparsity of very hot stars in this region, the high multiplicity of such stars (as our SED fitting code currently models all stars as single stars) or the high uncertainty on spectroscopic temperature of many of the O stars.

Fig. \ref{TempAur} also shows that our SED-fitted temperatures are in better agreement with the APOGEE spectroscopic temperatures than they are with the SIMBAD spectroscopic temperatures (which constitute most of the O-type stars). APOGEE spectra are generally more consistent and of better quality than the spectroscopy from SIMBAD, which might explain the difference. The median error on $T_{\rm eff}$ for the APOGEE spectroscopy is only $\sim$ 200 K, to be contrasted with $\sim$ 1100 K for the SIMBAD spectroscopic sample.  
\subsection{Clustering analysis with HDBSCAN}
\label{HDB}
In \citet{Quintana}, we identified kinematically-coherent OB associations in the Cygnus region by applying a flexible clustering method based on a Kolmogorov-Smirnov (KS) test on Galactic coordinates and proper motions. This choice was feasible because the OB associations were all at a similar distance. The distance spread of the OB stars in Auriga, on the other hand, is much larger.

For this work we therefore use the Hierarchical Density-Based Spatial Clustering of Applications with Noise (HDBSCAN, \citealt{HDBSCAN}) tool. It constitutes an extension of DBSCAN and identifies clusters by defining their cores through the number of neighbours within a radius $\epsilon$. In many clustering algorithms, including DBSCAN, the selection of clusters depends heavily upon the value of $\epsilon$. HDBSCAN overcomes this issue by allowing the user to define clusters at several density thresholds, therefore finding the most reliable groups and clusters.

In our testing, out of all HDBSCAN parameters, only \texttt{cluster\_selection\_method}, \texttt{min\_cluster\_size} and \texttt{min\_samples} were found to have an influence on the algorithm results. \texttt{Excess of mass (EOM)} and \texttt{Leaf} are the two selection methods. Whilst the former tends to identify larger structures and thereby decreases the noise (see e.g. \citealt{Kerr}), the latter outlines smaller and more homogeneous clusters, hence we favour this second choice as it is more suited to OB associations (see e.g. \citealt{Santos}). \texttt{min\_cluster\_size} sets the minimum number of stars for a cluster to be defined whereas \texttt{min\_samples} stands for the number of samples within a neighbourhood such that a point is treated like a core point \citep{HDBSCAN}. Varying \texttt{min\_cluster\_size} will only set which cluster is identified (i.e. a cluster is only identified if it has more members than \texttt{min\_cluster\_size}) while varying \texttt{min\_samples} will change the membership itself, and is therefore the most crucial parameter. 

The five parameters used for our clustering analysis are: $X$, $Y$, $Z$, $V_l$, $V_b$, where $XYZ$ are the Galactic cartesian coordinates and $V_l = 4.74 \ \mu_l \, \frac{d}{1000}$ is the transverse velocity in the $l$ direction in units of km s$^{-1}$ (with its equivalent in the $b$ direction). 

Our 5D parameter space thus contains three parameters in units of pc and two in km s$^{-1}$. Each parameter of the same units is normalised with respect to the parameter with the largest extent sharing this unit, i.e. $X$, $Y$ and $Z$ were normalised with respect to $X$ in order to overcome the stretching along the line-of-sight, while $V_b$ was normalised with respect to $V_l$. As such, all the normalised parameters have values between 0 and 1 but parameters with the same units are directly comparable.

To identify new OB associations we set \texttt{min\_cluster\_size} to 15 and \texttt{min\_samples} to 10, consistent with the typical minimum number of OB stars in OB associations \citep{Humphreys1978}. We apply HDBSCAN to the 5617 candidate OB stars with $\log(T_{\rm eff}) > 4$ or $log(\frac{L}{L_{\odot}}) > 2.5$. This threshold was chosen to include evolved high-mass stars with cooler temperatures \citep{Mamajek}, as we did in \citet{Quintana}. This process gave 14 groups listed in Table \ref{Aurassoc}.

Subsequently, based on the method by \citet{Santos}, we perform a bootstrapping process on the newly identified OB associations. We randomly vary the proper motions and the distance of each star within their uncertainties and apply HDBSCAN to the new sample. Each iteration gives us a new set of associations that we compare to the original associations. If a 'bootstrapped' associations has 5D parameters within 1$\sigma$ from the median of the original associations, then it corresponds to the same association. When this matching happens, we then compare the individual members of the bootstrapped association to the original association. We repeat this process 10,000 times, calculating the fraction of iterations in which a given association appears, and a fraction of those iterations in which a given star appears in that association. These fractions are taken as the probability that a given association is genuine and a membership probability for each star in each association. We also add stars that do not belong to the original associations, but appear in more than 50 \% of iterations in the bootstrapped associations. 

To estimate the reliability of our new associations we performed a Monte Carlo simulation to estimate how many OB associations, and with what properties, would be identified from a random distribution of stars. We randomly sampled the Galactic coordinates, PMs and SED-fitted distances of the 5617 candidate OB stars 100 times. For each iteration we ran HDBSCAN to identify new OB associations and performed the same bootstrapping process (with 1000 iterations) to estimate their probabilities. These simulations resulted in a total of 1154 'randomized' associations, i.e., an average of $\sim$12 per simulation. The probability for each of these associations is typically very low, with only 188 having probabilities greater than 50\%, 77 greater than 80\% and 46 greater than 90\%, equivalent to $\sim$2, $\sim$1 and $<$1, on average, per simulation. In the real data, we identified 9 groups with probabilities $>$50\%, 7 with probabilities $>$80\% and 4 with probabilities $>$90\%. Comparison with our simulation suggests that the 4 associations with probabilities $>$90\% are likely to all be real (especially since their probabilities are all $>$99\%), while the 5 associations with probabilities of 50--90\% may include 2 contaminants.

Our simulations do show that false-positive, high-probability associations ($>$80\%) are almost entirely found nearby ($d \lesssim 1.5$ kpc). We therefore discard all the nearby ($<$1.5 kpc) OB associations with a probability lower than 90 \%, retaining only the very high probability groups (now named associations 1-4) and the very distant group with a moderately-high probability (association 5). The 4 discarded candidate associations would require further data (e.g. RVs), more precise astrometry or expanded membership amongst lower-mass stars to confirm them as being real.

\begin{table}
	\centering
	\caption{Properties of the newly-identified OB associations in Auriga. $N$ is the initial number of stars in the association (before bootstrapping), $N_{\rm g}$ is the number of likely members (with a membership probability of at least 50 \%) and $N_{\rm tot}$ is the total number of stars in the associations, adding those appearing during the bootstrapping with a probability of at least 50 \%. $d_m$ stands for the median distance of the group. Probability gives the probability that the association is real. \label{Aurassoc}}
	\renewcommand{\arraystretch}{1.3} 
	\begin{tabular}{lccccccccr} 
		\hline
		Association & $N$ & $N_{\rm g}$ & $N_{\rm tot}$ & $d_m$ (pc) & Probability(\%) \\
		\hline
		 & 26 & 20 & 21 & 738 & 86.58  \\
		 & 18 & 18 & 25  & 906 & 83.20 \\
		1 & 198 & 186 & 215 & 1056 & 99.98 \\
		2 & 41 & 41 & 43 & 1085 & 99.99 \\
		 & 17 & 16 & 16 & 1475 & 57.56 \\
		3 & 99 & 89 &  119 & 1514 & 99.93\\
		4 & 130 & 127 & 138 & 1923 & 99.87 \\
		 & 15 & 13 & 13 & 1956 & 4.15 \\
		 & 37 & 19 & 19 & 2188 & 48.99 \\
		 & 23 & 11 & 11 & 2508 & 9.35 \\
		 & 21 & 9 & 9 & 2677 &  18.24 \\
		5 & 90 & 39 & 39 & 2760 & 82.31 \\
		 & 83 & 13 & 13 & 2803 & 63.10 \\
		 & 15 & 7 & 7 & 2951 & 8.35\\	
		\hline
	\end{tabular}
\end{table}

\begin{figure*}
    \centering
    \includegraphics[scale =0.45]{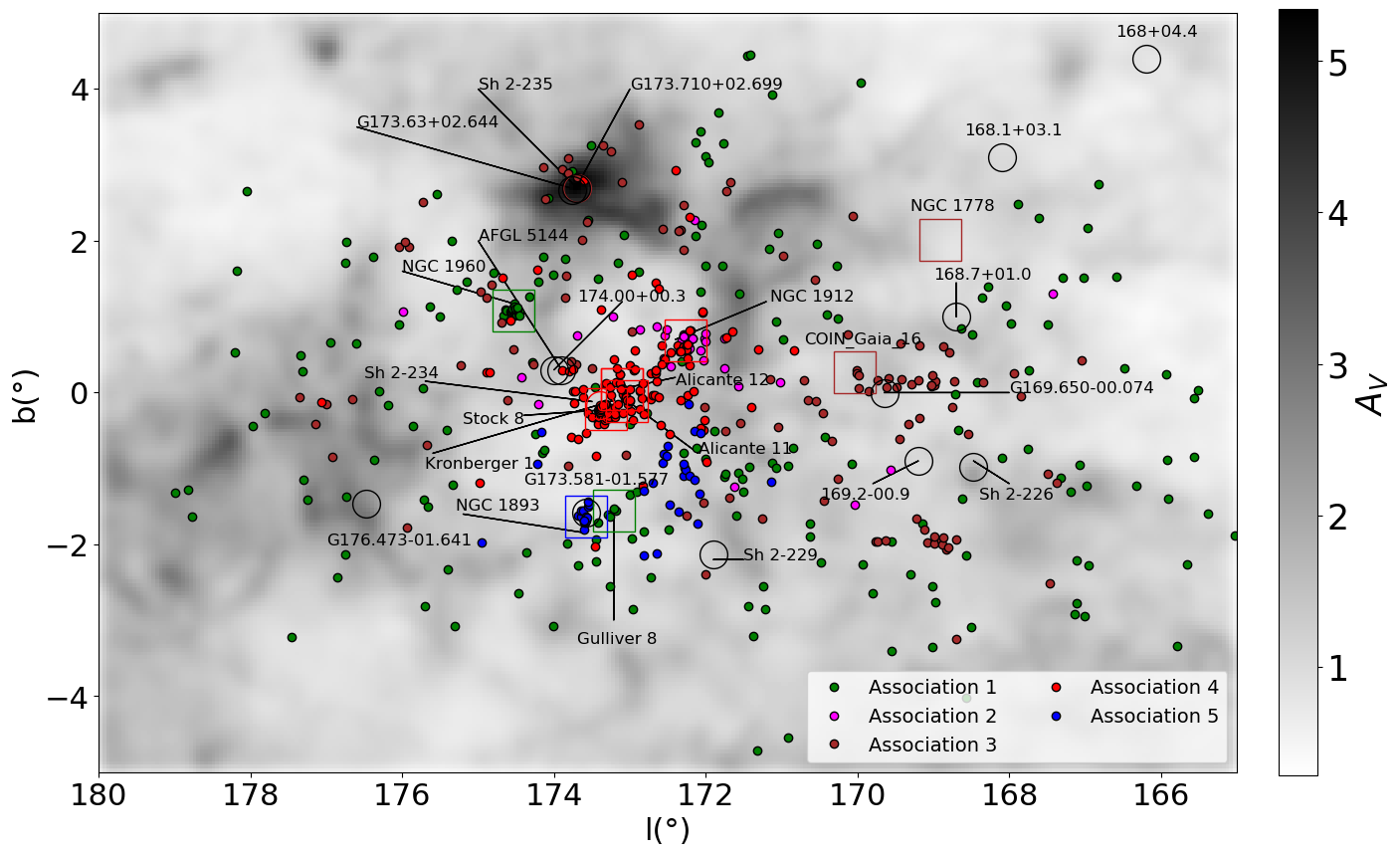}
    \caption{Spatial distribution in Galactic coordinates of the 5 new OB associations in Auriga. The background extinction map and the features highlighted in this map are the same as in Fig. \ref{AurSpace}.}
    \label{GalAurAssoc}
\end{figure*}
\begin{figure}
    \centering
    \includegraphics[width = \columnwidth]{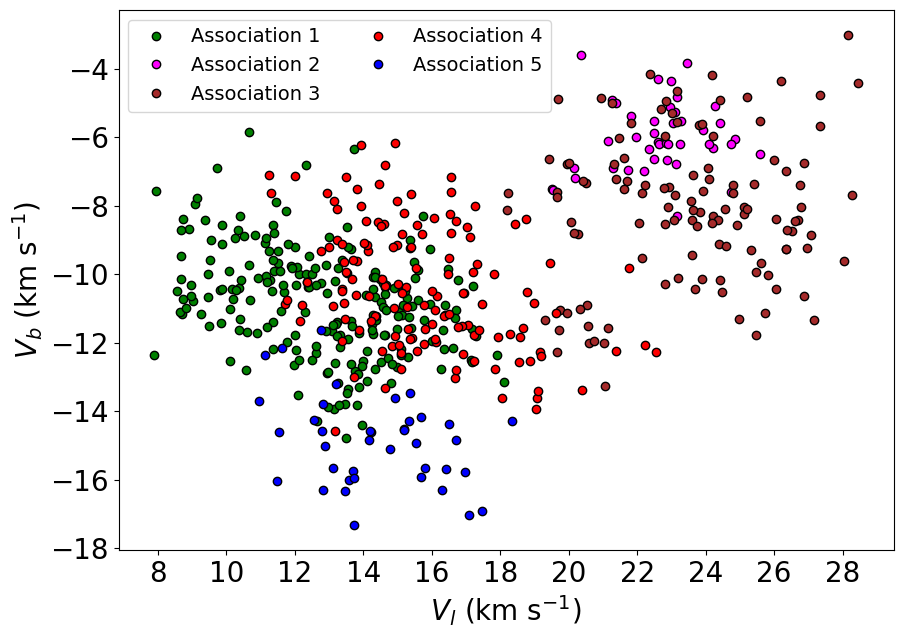}
    \caption{Transverse velocity distribution of the 5 new OB associations in Auriga.}
    \label{VelAurAssoc}
\end{figure}
\begin{figure*}
    \centering
    \includegraphics[scale = 0.45]{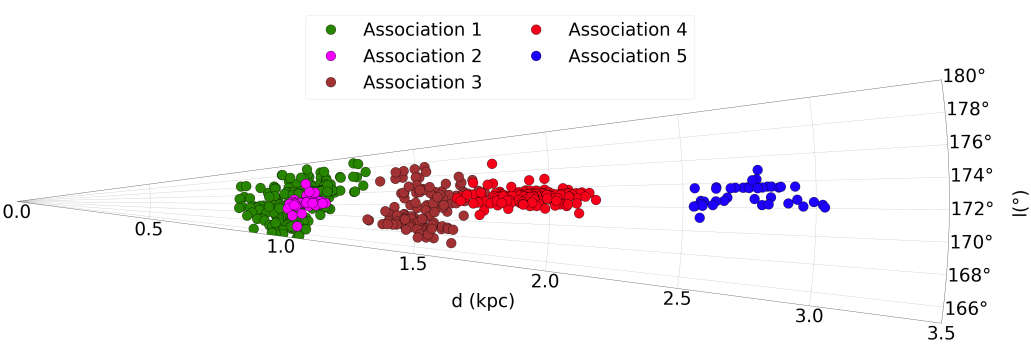}
    \caption{Galactic longitude as a function of SED-fitted distance for the 5 new OB associations in Auriga. The median error bars on distances are respectively $\sim$ 30 pc for associations 1 and 2, $\sim$ 50 pc for association 3, $\sim$ 70 pc for association 4 and $\sim$ 160 pc for association 5.}
    \label{DistAurAssoc}
\end{figure*}

The result of this process is that we are left with 5 new high-confidence, spatially- and kinematically-coherent OB associations in the Auriga region. We show them in Galactic coordinates in Fig. \ref{GalAurAssoc}, in Galactic transverse velocity in Fig. \ref{VelAurAssoc} and in distance in Fig. \ref{DistAurAssoc}. These new OB associations are distributed over a range of distances from 1 kpc to almost 3 kpc, with many super-imposed on each other on the plane of the sky, explaining the difficulty separating their members with pre-\textit{Gaia} data.

\subsection{Comparison with historical associations and open clusters}
\label{Comphistoc}
We crossmatch the members of our new OB associations with the historical members of Aur OB1 and OB2 from \citet{Melnik2020} and the open cluster members from \citet{CantatGaudin}, with the results displayed in Table \ref{AurComp}.
\begin{table}
	\centering
	\caption{Comparison between our new OB association members and OB stars in the historical associations and in the open clusters from \citet{CantatGaudin}. $N_{\rm hist}$ stands for the number of stars in a historical association whilst $N_{\rm OC}$ designates the number of stars in an open cluster. The notations CG16, G8, K1 and S8 stand respectively for COIN-Gaia\_16, Gulliver 8, Kronberger 1 and Stock 8. \label{AurComp}}
	\renewcommand{\arraystretch}{1.3} 
	\begin{tabular}{lccccccccr} 
		\hline
		Assoc. & $N_{\rm hist}$ & Hist. assoc. & $N_{\rm OC}$ & OC  \\
		\hline
		1 & 7 & Aur OB1 & 25, 4 & NGC 1960, G8 \\
		2 & & & 26 & NGC 1912 \\
		3 & 1 & Aur OB1 & 9, 4 & NGC 1778, CG16 \\
		4 & 1, 8 & Aur OB1, OB2 & 49, 5 & S8, K1 \\
		5 & 2 & Aur OB2 & 15 & NGC 1893 \\
		\hline
	\end{tabular}
\end{table}

Association 1 includes stars in both Aur OB1 and NGC 1960 and is the largest foreground association in the area. The other historical members of Aur OB1 are spread over the other new OB associations. Associations 4 and 5 have significant overlaps with Stock 8 and NGC 1893. This comparison suggests that NGC 1893 is located closer than previous estimations \citep{Lim2018,CantatGaudin} at a distance of $\sim$2.8 kpc, consistent with the distance from \citet{Melnik2009}.
\section{Analysis of the new OB associations}
\label{analysis}
In this section we perform a kinematic and physical analysis of the new OB associations in Auriga, studying their expansion and star formation history.

\begin{table*}
	\centering
	\caption{Properties of the new OB associations. The first column indicates the parameter, where the subscript 'm' indicates the median value and '$\sigma$' the dispersion. The total initial stellar mass is corrected for observational incompleteness, as described in the text.  \label{TabParamGroups}}
	\renewcommand{\arraystretch}{1.5} 
	\begin{tabular}{lccccccr} 
		\hline
		Parameters & Units & Assoc. 1 & Assoc. 2 & Assoc. 3 & Assoc. 4 & Assoc. 5 \\
		\hline
        ${\rm RA(ICRS)}_m$ & deg & 81.30 & 82.13 & 80.16 & 82.02 & 80.70  \\
        ${\rm DE(ICRS)}_m$ & deg & 34.97 & 35.84  & 36.57 & 34.77 & 33.94  \\
        $l_m$  & deg & 170.72 & 172.24  & 170.70 & 173.09 & 172.82  \\
        $b_m$  & deg & -0.16 & 0.70  & 0.11 & -0.03 & -1.48  \\
        $d_m$ & pc & 1056 & 1085  & 1514 & 1923 & 2760 \\
        $\sigma_{d}$ & pc & 102.2 & 25.2  & 76.0 & 103.8 & -  \\
        $V_{l_m}$ & km s$^{-1}$ & 12.98 & 22.85  & 23.69 & 15.37 & 14.18  \\
        $\sigma_{V_l}$ & km s$^{-1}$ & 2.52 & 1.09  & 2.96 & 2.20 & 2.10  \\
        $V_{b_m}$ & km s$^{-1}$ & -10.75 & -6.16  & -8.10 & -10.58 & -14.84  \\
        $\sigma_{V_b}$ & km s$^{-1}$ & 1.41 & 0.95  & 2.00 & 2.05 & 1.17 \\
        Observed number of B stars & & $194 \pm 3$ &  $40^{+2}_{-1}$  &  $107^{+3}_{-2}$ & $115^{+3}_{-4}$ &  $32 \pm 2$ \\
        Observed number of O stars & & $12^{+3}_{-2}$ & $0 \pm 0$ & $4 \pm 1$ & $13 \pm 2$ & $3^{+2}_{-1}$ \\
        Total stellar initial mass & $M_{\odot}$  & $6051^{+426}_{-387}$ & $1219^{+182}_{-167}$ & $3075^{+298}_{-276}$ & $3500^{+315}_{-306}$ &  $879^{+163}_{-136}$  \\
        HR diagrams age & Myr  & 0-30 & -  & 0-20 & 0-5 & 0-10  \\
        Related OCs age &  Myr &  18-26 & 250-375  & 26 & 4-8 & 1-5 \\
        Traceback age & Myr &  $20.9^{+1.1}_{-1.2}$ & $369.9^{+8.3}_{-22.2}$  & $11.7^{+7.2}_{-3.0}$ & $1.6^{+1.3}_{-0.9}$ & - \\
        Age & Myr & $\sim$ 20 & - & 10-20 & 0-5 & 0-10  \\
		\hline
	\end{tabular}
\end{table*}
\subsection{Physical properties of the individual associations}
We have estimated the observed number of O- and B-type stars in each association. To do so we defined B-type stars as those with SED-fitted $\log(T_{\rm eff}) > 4$ and $\log(T_{\rm eff}) < 4.3$ and O-type stars as those with SED-fitted $\log(T_{\rm eff}) > 4.3$, using the same thresholds than in Section \ref{genresults}. Uncertainties were estimated through a Monte Carlo experiment where the effective temperature of each star was randomly sampled within their uncertainties. This is shown in Table \ref{TabParamGroups} and, in line with the HR diagrams in Fig. \ref{HRAur}, shows a dominance of late B-type stars and a few O-type stars.

To determine the total mass of each association, we first identified the range of masses over which our sample completeness is expected to be unbiased by the age of our stars. We chose this mass range to be 2.5 $M_{\odot}$ (the mass of an A0 star) to 7.1 $M_{\odot}$ (the post main-sequence turn-off mass at an age of 50 Myr, \citealt{Ekstrom}). We then corrected the number of stars according to the incompleteness levels we have calculated and displayed in Fig. \ref{Incompleteness}.

We performed a Monte Carlo simulation sampling stellar masses at random using the mass function from \citet{IMFMasch}. We counted both the number of stars in our selected mass range and the total number and mass of stars. We stopped the simulation only when we reached the total number of observed stars in the selected mass range. This process was repeated 10,000 times, using the uncertainties on the individual SED-fitted stellar masses, to obtain an uncertainty for the total stellar mass of each association. These are provided in Table \ref{TabParamGroups} and range from $\sim$900 to $\sim$6000 M$_\odot$. The most massive is association 1, with an estimated initial stellar mass of $\sim$6000 M$_\odot$ and currently containing about 200 B-type members.

\subsection{Kinematic properties of the individual associations}
We calculated the median coordinates (equatorial and galactic), distances and transverse velocities for each OB association. In addition, we have computed the intrinsic dispersion in distance and transverse velocities based on the method from \citet{Ivezic} using the observational uncertainties. The distance dispersions typically range up to a few tens of pc, while the velocity dispersions range up to 3 km s$^{-1}$, consistent with that of other OB associations \citep{Wright2020}.
\subsection{HR diagrams of association members}
\begin{figure*}
    \centering
    \includegraphics[scale = 0.15]{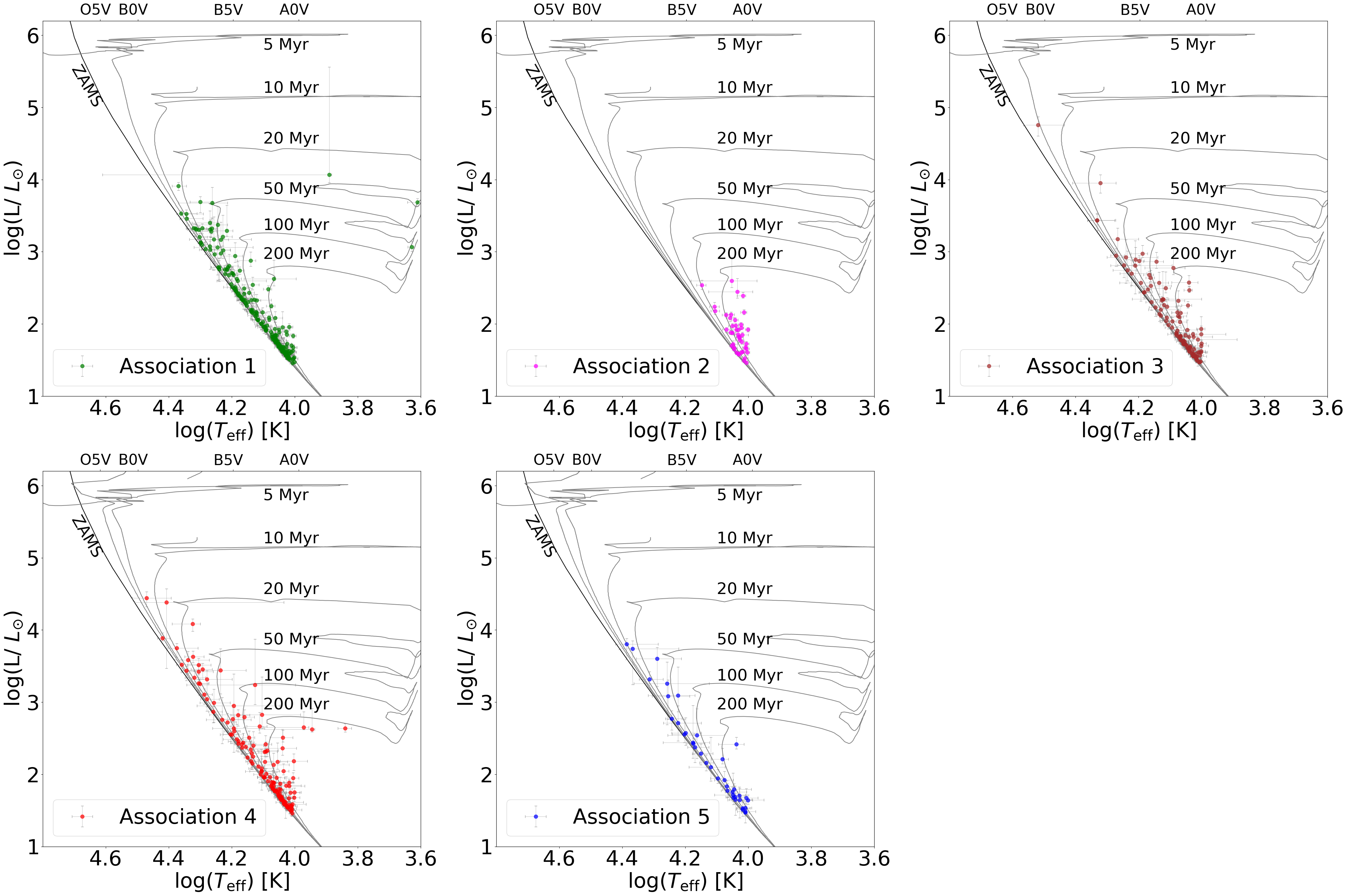}
    \caption{HR diagrams for the members of the new OB associations in Auriga. Isochrones have been shown from the rotating evolutionary models from \citet{Ekstrom}. Positions of some spectral types have been indicated on the top horizontal axis for clarity.}
    \label{HRAur}
\end{figure*}
Fig. \ref{HRAur} shows the HR diagrams for each association, produced with the SED-fitted effective temperatures and luminosities. It is clear that the identified members are dominated by late B-type stars, preventing a straightforward age assessment. Association 1 contain a few stars close to the 50, 100 and 200 Myr isochrones, which would make this much older than other known OB associations, though these may be contaminants. Associations 3 to 5 contain hotter stars that suggest a younger age of $<$ 20 Myr. 

There are a number of factors that can affect the positions of stars in the HR diagram. First among these is extinction, which is derived for each star individually as part of our SED fitting process from the extinction map. The uncertainty in the extinction to a given star, which is derived from the distance uncertainty, propagates through to the uncertainty on the effective temperature and luminosity shown in the HR diagram. An alternative approach might be to use a single extinction value for all members of an association, but the effect of this will be small as the variation in extinction across members of an association is small, with a typical standard deviation in $A_V$ of 0.2 - 0.6 mag. Such a difference in reddening does have a small effect on the derived physical parameters. However, if we reproduce these HR diagrams using the median extinction for all association members, while there are small changes in the position of each star, the extreme outliers do not change significantly.

More significant factors affecting the position of stars in the HR diagram include binarity, and the presence of possible contaminants. Association 1 constitutes a good example of this, for which most of its stars are close to the ZAMS and therefore consistent with being under 20 Myr old, suggesting that its stars sitting on the 50, 100 and 200 Myr isochrones may be contaminants.
\subsection{Expansion and traceback age}
\label{traceback}
Fig. \ref{HRAur} shows that many of the OB associations in Auriga have ages of several tens of Myr. Therefore, instead of using a linear expansion model to determine their expansion age, we trace back the associations using the epicycle approximation from \citet{Fuchs}, and correct for the local standard of rest (LSR) from the values of \citet{Scho}. We gather RVs from the APOGEE survey and from SIMBAD. RVs from the literature are discarded if they lack an uncertainty or if their measurement is considered unreliable. If some stars belong to both APOGEE and SIMBAD, we take the weighted mean of the two values. In doing so we obtain a sample of 95 stars with RVs. 

We calculated the median RV for each association and track back whole associations rather than individual stars, due to the effects of unresolved close binaries. Again we apply a Monte Carlo's simulation to compute the uncertainties on the median velocities following the method in \citet{Quintana2022}. The results are shown in Table \ref{Radvel}.

\begin{table}
	\centering
	\caption{RVs calculated for the new OB associations. $N_{RV}$ is the number of stars with a reliable measured RV. References are: (1): APOGEE, (2): \citet{Fehren1992}, (3): \citet{Grenier1999}, (4): \citet{Gont}, (5): \citet{Turner2011}, (6): \citet{Choj}, (7): \citet{GaiaDR2}, (8): \citet{Zhong2020}  \label{Radvel}}
	\renewcommand{\arraystretch}{1.3} 
	\begin{tabular}{lccccccccr} 
		\hline
		Assoc. & $N_{RV}$ & RV (km s$^{-1}$) & References \\
		\hline
        1 & 47 & $2.13 \pm 1.26$ & All but (6)  \\
        2 & 5 & $6.59 \pm 4.31$ & (8) \\
        3 & 8 & $-4.64 \pm 3.01$ & (1), (6), (8) \\
        4 & 31 & $4.12 \pm 1.63$ & (1), (7), (8)   \\
        5 & 4 & $1.46 \pm 7.45$ & (8) \\
		\hline
	\end{tabular}
\end{table}

\begin{figure}
    \centering
    \includegraphics[scale = 0.3]{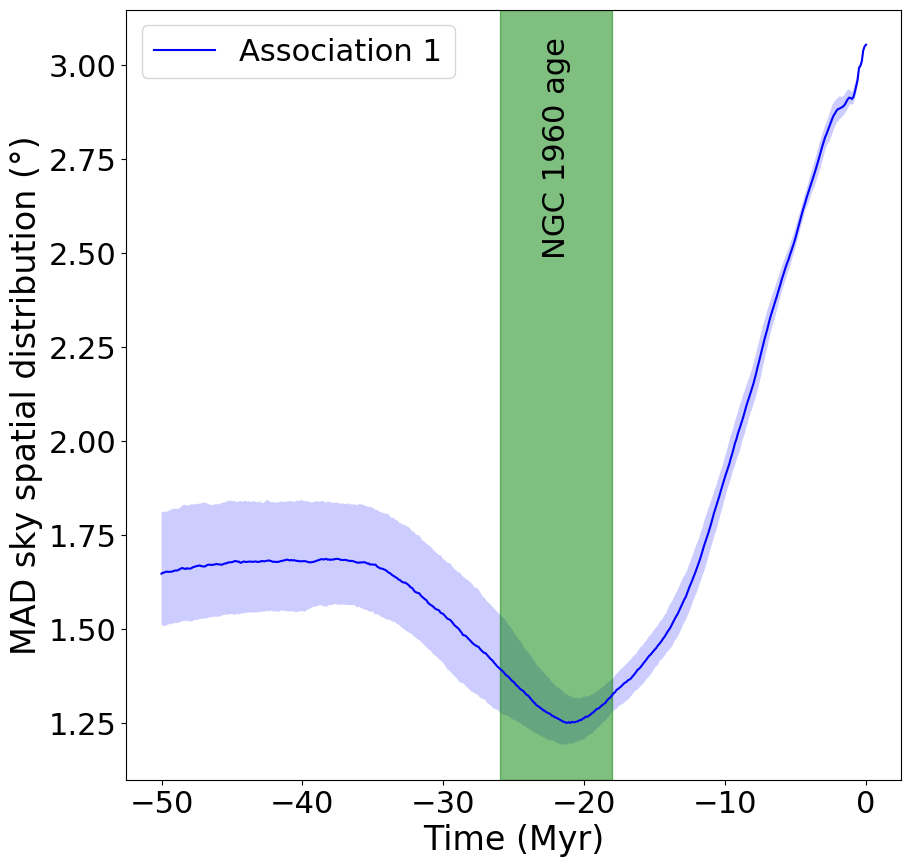}
    \caption{MAD of the on-sky spatial distribution for members of association 1 as a function of traceback time. The age of the related open cluster (NGC 1960) is shown, taken from Table \ref{AurOCs}.}
    \label{TracebackAssoc3}
\end{figure}

Combining RVs with Gaia PMs allows us to perform a 3D traceback on these associations. We use the HR diagrams (Fig. \ref{HRAur}) together with the ages of the related open clusters to add constraints on the age estimations, which are displayed in Table \ref{AurOCs}. With this information, we set the upper limit on traceback to 50 Myr in the past for associations 1 and 5, 400 Myr for association 2, 30 Myr for association 3 and 20 Myr for association 4. We trace back each association in time steps of 0.1 Myr, and at each time step we calculate the MAD (median absolute deviation) in Galactic coordinates of the on-sky spatial distribution of association members\footnote{We do not calculate the MAD in 3D due to the large error bars on RVs that causes uncertain line-of-sight distances as we go back further in the past.}, and we estimate the time of minimum MAD when the association is at its most compact. We repeat this process 1000 times to derive uncertainties. An example is shown in Fig. \ref{TracebackAssoc3} for association 1, while the other associations are shown in Figure \ref{TracebackHR}.

We estimated ages for each association based on the combination of (i) the time for the system to trace back to its most compact state, (ii) the age of any open cluster or star-forming region linked to the association (Section \ref{Comphistoc}), and (iii) any age constraints arising from the HR diagram (Fig. \ref{HRAur}). These ages are listed in Table \ref{TabParamGroups}. For some systems the traceback was able to place reasonable constraints on the age of the system, (e.g., for association 1), while for other associations the best constraint came from the open clusters and star-forming regions the association was linked to (e.g., for associations 4 and 5). For the remaining associations either the HR diagram provided the best constraint on the age (e.g., for association 3) or very little constraint was possible by any means (e.g., association 2).

\section{Discussion}
\label{discussion}
In this section we discuss our findings and how the new Auriga OB associations can help understand the star formation history of the region. Our main results are outlined as follows:
\begin{itemize}
   \item We have shown that both Aur OB1 and Aur OB2 are too extended in PM and distance to be genuine OB associations, encouraging us to revisit the census of OB stars and associations in the region.
    \item We identified more than 5000 candidate OB stars across the region using our SED fitter, with an estimated reliability of 90 \%.
   \item We identified 5 new high-confidence OB associations in the area that we analysed physically and kinematically, and estimated their age through a combination of 3D kinematic traceback, their link to open clusters and star-forming regions with known ages, and the distribution of members in the HR diagram. Only a small fraction ($\sim$10 \%) of the identified OB stars have been assigned to these associations.
\end{itemize}
\subsection{The new Auriga OB associations}
We have identified 5 new OB associations in Auriga, with total stellar masses from a few hundreds to a few thousand solar masses, and with kinematic properties consistent with other OB associations \citep{Wright2020}. They are likely related to open clusters in the area (see Table \ref{AurComp}). 

Association 1 shares several members with Aur OB1 and is related to NGC 1960, so it should be seen as the replacement for the historical Aur OB1 association. Similarly, the historical members of Aur OB2 are now divided between associations 4 and 5, respectively related to Stock 8 and NGC 1893. This confirms the suggestion of \citet{Marco} to divide Aur OB2 into two different associations, one in the foreground and one in the background. 

The H{\sc ii} region Sh 2-235 instead appears to be related to association 3, since it is located at a similar distance of $1.36 \pm 0.27$ kpc \citep{Foster2015}. Association 3 also includes HD 36483, an O9.5IV star \citep{GOSS}, which may be responsible for ionizing the H{\sc ii} region.

Association 4 occupies the centre of our region of study, where three OCs are found, along with the  H{\sc ii} region Sh 2-234 (Fig. \ref{GalAurAssoc}). Sh 2-234 is located at a distance of $2.19 \pm 0.10$ kpc \citep{Foster2015}. Its relation to Aur OB2 and the surrounding OCs has been suggested by \citet{Marco} and we confirm it to be related to association 4. We cannot comment on whether the star LS V +34 23 is part of association 4 (previously in Aur OB2 and thought to be responsible for ionizing Sh 2-234, \citealt{Marco}) as its \textit{Gaia} photometry failed our quality checks, preventing us from performing an SED fit. However, association 4 does contain LS V + 34 15, LS V + 34 21 and LS V + 35 25, of respective spectral types O5.5V \citep{Negueruela2007}, O9IV \citep{RomanLopes2019} and O9.5V \citep{Georgelin1973}, each probable sources of ionization for the H{\sc ii} region Sh 2-234. However, the RV of Sh 2-234 has been measured as $-21.4 \pm 0.2$ km s$^{-1}$ \citep{Anderson2015}, which is significantly different from the RV we estimated for association 4 (Table \ref{Radvel}), even if those stars are responsible for ionizing the  H{\sc ii} region, the association and the  H{\sc ii} region may not otherwise be related.
\subsection{Expansion and age of the OB associations}
Our analysis revealed that our OB associations have various ages, from the youngest associations 4 and 5 with ages $<$ 10 Myr to associations of several tens of Myr old (association 3). For the OB associations where multiple age indicators were available, the ages derived by different methods were consistent. The exception to this is association 2, with the majority of its OB stars consistent with being on the ZAMS (Fig. \ref{HRAur}) while its related OC is several hundreds of Myr old (Table \ref{AurOCs}), far older than most OB associations.

In Section \ref{traceback} we showed that nearly all our OB associations traced back into a more compact configuration in the past, which is a signature of expansion (see e.g. \citealt{WrightMamajek} and \citealt{MiretRoig2022}). We however point out that associations 4 and 5 reached their most compact state very recently (Fig. \ref{TracebackHR}).
\subsection{OB stars unassigned to groups}
The 5 OB associations contain 554 OB stars in total from our sample of 5617 SED-fitted OB stars in the area. This means that $\sim$90 \% of the OB stars have been unassigned to any stellar group, which could be explained by several factors.

In Section \ref{HDB}, we imposed a minimum size of 15 OB stars per association to be consistent with their definition \citep{Humphreys1978,Wright2020}. There could be other stellar groups in the area which are dominated by low-mass stars and only contain a handful of OB stars. Similarly, some OB stars initially belonging to a group were rejected during the bootstrapping process (see Table \ref{Aurassoc}). This was particularly the case for most distant stars ($>2$ kpc) as the \textit{Gaia} parallaxes become less precise.

Our sample includes many late B-type stars. A B9 star with an initial stellar mass of 2.75 $M_{\odot}$ \citep{Mamajek} has a lifetime of $\sim$ 700 Myr as predicted by stellar evolutionary models \citep{Ekstrom}. This value is far beyond the typical lifetime of an OB association \citep{Wright2020} and implies that even if those stars were born clustered, they would probably have dispersed into the Galactic field population since.

It is also possible that some of these OB stars formed within associations but were ejected and became runaways. Notably, massive stars are more likely to belong to multiple systems \citep{Lada2006}, and when the primary star undergoes a supernova explosion, it can eject the secondary star beyond the group it was born into.
\subsection{Distribution of OB associations and Galactic structure}
\label{spiralarms}
\begin{figure*}
    \centering
    \includegraphics[scale = 0.28]{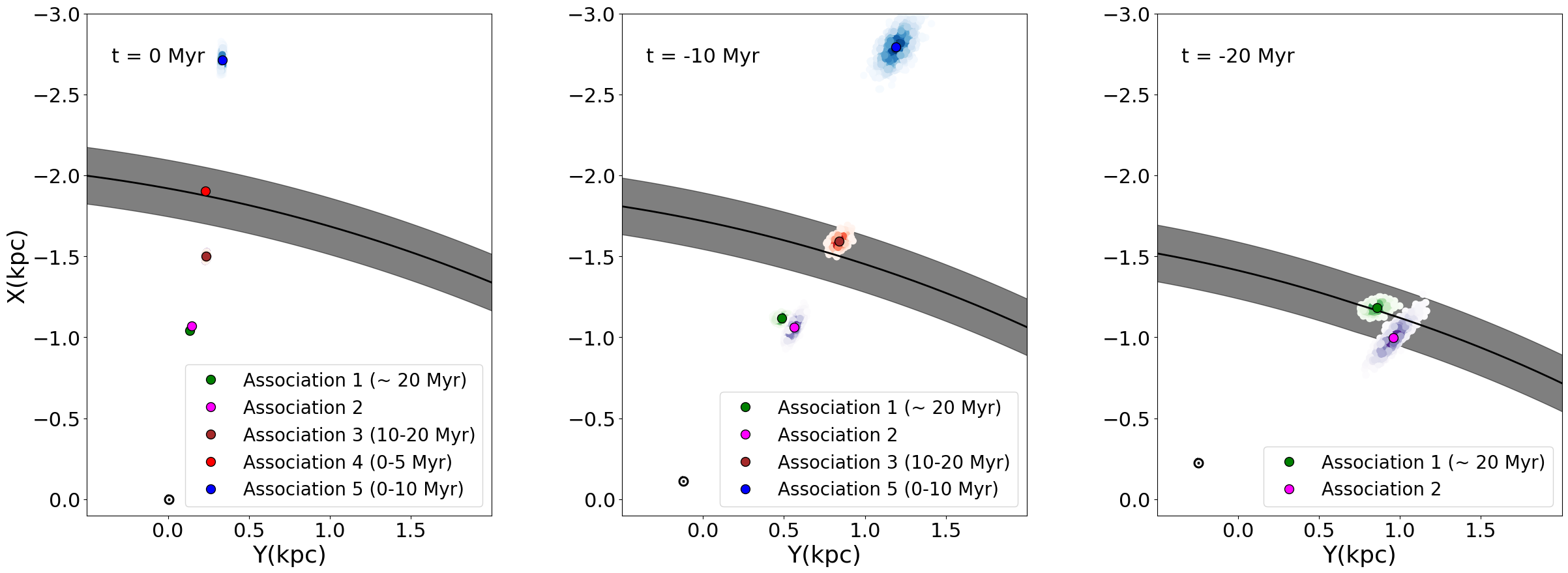}
    \caption{Median positions of the OB associations across the X-Y plane from the present time to 20 Myr in the past, shown relative to the Local Standard of Rest. Uncertainties on their position from the traceback have been shown as a blur. The Perseus spiral arm model, including its thickness, is from \citet{Reid2019}. The Sun symbol corresponds to the position of the Sun.}
    \label{AurSpiralArms}
\end{figure*}
The Perseus spiral arm intercepts our sightline at a distance of approximately 2 kpc \citep{Reid2019}, at approximately the position of association 4, the youngest of our new OB associations. Fig. \ref{AurSpiralArms} shows the positions of our new OB associations, with their ages, relative to the position of the Perseus spiral arm. While association 4 is coincident with the current position of the Perseus spiral arm, the associations closer to us are older, indicating a potential age gradient.

To determine whether this age gradient is related to the motion of the Perseus spiral arm, we model the positions of the spiral arm and our new OB associations over the last 20 Myrs. We use the spiral arm model from \citet{Reid2019} and the spiral arm pattern speed of $\Omega_p = -28.2 \pm 2.1$ km s$^{-1}$ from \citet{Dias2019}. We trace back the position of the Perseus spiral arm 20 Myr into the past in the frame of the LSR. 
 
 Fig. \ref{AurSpiralArms} shows the position of the OB associations with the Perseus spiral arms, at intervals of 10 Myr, back to 20 Myr in the past. At 10 Myr in the past it is clear that association 3 (with an estimated age of 10-20 Myr) is coincident with the spiral arm, while at 20 Myr in the past, association 1 (estimated age of $\sim$20 Myr) is coincident with the spiral arm. Association 2 crosses the spiral arm $\sim$20 Myr ago as well, despite its related OC and traceback suggesting an older age (Fig. \ref{TracebackHR}), which may suggest that the association is younger and not related to NGC 1912, or that the association did not form within the spiral arm. As for association 5, it stays too distant to be related to the Perseus spiral arm and may have formed outside (or in the Outer spiral arm).
 
This result shows that OB associations can be used not only as tracers for the current positions of spiral arms but also as a probe for the star formation history of a region and potentially the progress of a spiral arm through the region.

\section{Conclusions}
\label{conc}
We have shown that Aur OB1 and OB2 are not genuine OB associations, because their members are characterized by a too large spread in proper motion and distance. Applying an improved SED fitting tool, we have identified 5617 OB stars with a reliability of $\sim$ 90 \% for the lowest temperature threshold. 

Using a clustering algorithm (HDBSCAN), we have identified 5 high-confidence OB associations that we connect to the open clusters and star-forming regions in the area. Association 1 is the main foreground association at a distance of $\sim$ 1 kpc and with a mass of $\sim$ 6000 $M_{\odot}$ and should replace Aur OB1 due to its common members and relation with NGC 1960. Similarly, we argue that Aur OB2 should be replaced by association 4 (at $\sim$ 1.9 kpc and with a total mass of $\sim$ 3500 $M_{\odot}$), and 5 (at $\sim$ 2.8 kpc and with a total mass of $\sim$ 900 $M_{\odot}$).

We have analysed these OB associations, combining HR diagrams and kinematic traceback, to constrain their ages. We have also studied their expansion, their total stellar masses, their number of OB stars and their 3D position.

We have also identified an age progression between several of these associations, that suggests their origins may have been within the Perseus spiral arm. This shows that OB associations constitute useful tools to study recent star formation history and the position and motion of the Galactic spiral arms.
\section*{Acknowledgements}
We thank the anonymous referee for their thorough reading of the manuscript and their insightful comments. ALQ acknowledges receipt of an STFC postgraduate studentship. NJW acknowledges receipt of a Leverhulme Trust Research Project Grant (RPG-2019-379). The authors would like to thank John Southworth for discussions, and Eleonara Zari for her help to calculate the position of the spiral arms.

This paper makes uses of data processed by the Gaia Data Processing and Analysis Consortium (DPAC, https://www.cosmos.esa.int/web/gaia/dpac/consortium) and obtained by the Gaia mission from the European Space Agency (ESA) (https://www.cosmos.esa.int/gaia), as well as the INT Galactic Plane Survey (IGAPS) from the Isaac Newton Telescope (INT) operated in the Spanish Observatorio del Roque de los Muchachos. Data were also based on the Two Micron All Star Survey, which is a combined mission of the Infrared Processing and Analysis Center/California Institute of Technology and the University of Massachusetts, along with The UKIDSS Galactic Plane Survey (GPS), a survey carried out by the UKIDSS consortium with the Wide Field Camera performing on the United Kingdom Infrared Telescope.

This work also benefited from the use of \textit{TOPCAT} \citep{Topcat}, Astropy \citep{Astropy} and the Vizier and SIMBAD database, both operated at CDS, Strasbourg, France.


\section*{Data Availability}
The data underlying this article will be uploaded to Vizier.



\bibliographystyle{mnras}
\bibliography{Bibliography} 




\appendix

\section{Traceback diagrams}
\label{indiv}
\begin{figure*}
\centering
\includegraphics[scale = 0.18]{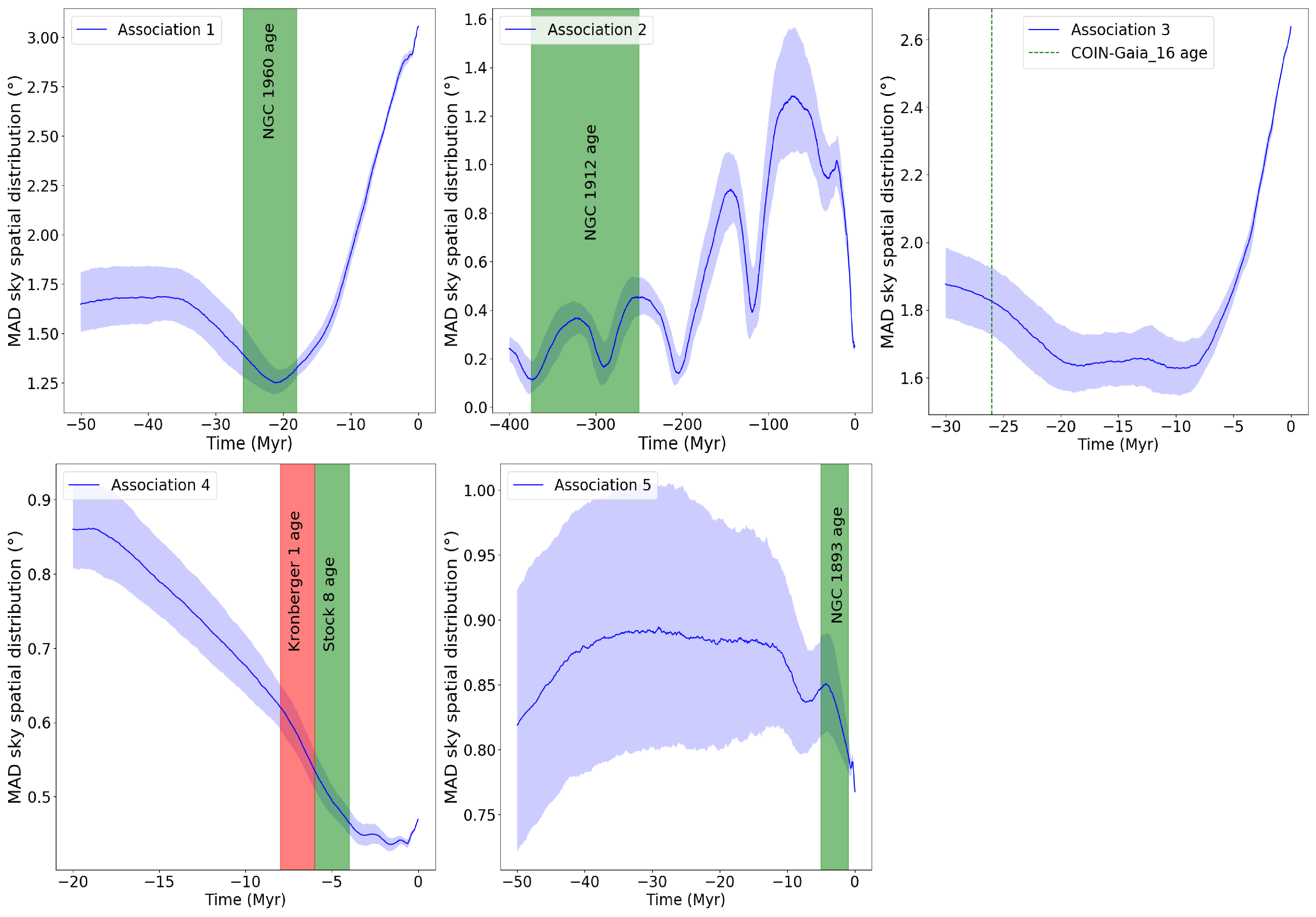}
\caption{MAD of the on-sky distribution of members of each association as a function of traceback time. \label{TracebackHR}}
\end{figure*}

\bsp	
\label{lastpage}
\end{document}